\documentclass[prd,tightenlines,amsmath,amssymb,showpacs,superscriptaddress,nofootinbib]{revtex4}
\usepackage{graphicx,epsfig}
\usepackage{dcolumn}
\usepackage{bm}
\usepackage{rotating}
\usepackage{color}
\usepackage{subfigure}
\usepackage{pstricks}
\usepackage{pst-node}
\usepackage{ifpdf}
\usepackage{overpic}

\begin{document}

\normalsize
\parskip=5pt plus 1pt minus 1pt

\title{\boldmath Observation of $\eta_{c}$ decay into $\Sigma^{+}\bar{\Sigma}^{-}$ and $\Xi^{-}\bar{\Xi}^{+}$ final states}

\author{\small
M.~Ablikim$^{1}$, M.~N.~Achasov$^{5}$, O.~Albayrak$^{3}$, D.~J.~Ambrose$^{39}$, F.~F.~An$^{1}$, Q.~An$^{40}$, J.~Z.~Bai$^{1}$, Y.~Ban$^{27}$, J.~Becker$^{2}$, J.~V.~Bennett$^{17}$
% ...
\\
\vspace{0.2cm}
(BESIII Collaboration)\\
\vspace{0.2cm} {\it
$^{1}$ Institute of High Energy Physics, Beijing 100049, P. R. China\\
$^{2}$ Bochum Ruhr-University, 44780 Bochum, Germany\\
$^{3}$ Carnegie Mellon University, Pittsburgh, PA 15213, USA\\
% ...
$^{5}$ G.I. Budker Institute of Nuclear Physics SB RAS (BINP), Novosibirsk 630090, Russia\\
% ...
$^{17}$ Indiana University, Bloomington, Indiana 47405, USA\\
% ...
$^{27}$ Peking University, Beijing 100871, P. R. China\\
% ...
$^{39}$ University of Rochester, Rochester, New York 14627, USA\\
$^{40}$ University of Science and Technology of China, Hefei 230026, P. R. China\\
% ...
}
}

\begin{abstract}

  Using a data sample of $2.25\times10^{8}$ $J/\psi$ events collected
  with the BESIII detector, we present the first observation of the
  decays of $\eta_{c}$ mesons to $\Sigma^{+}\bar{\Sigma}^{-}$ and
  $\Xi^{-}\bar{\Xi}^{+}$. The branching fractions are
  measured to be $(2.11\pm0.28_{\rm stat.}\pm0.18_{\rm
    syst.}\pm0.50_{\rm PDG})\times10^{-3}$ and $(0.89\pm0.16_{\rm
    stat.}\pm0.08_{\rm syst.}\pm0.21_{\rm PDG})\times10^{-3}$ for
  $\eta_{c} \to \Sigma^{+}\bar{\Sigma}^{-}$ and
  $\Xi^{-}\bar{\Xi}^{+}$, respectively.  These branching fractions
  provide important information on the helicity selection rule in
  charmonium-decay processes.

\end{abstract}


\maketitle

\section{\boldmath{Introduction}}

Experimental studies on exclusive charmonium decays play an important
role in testing perturbative Quantum Chromodynamics (pQCD). In the
Standard Model (SM), the $\eta_{c}$ meson is the lowest lying
charmonium state in a 0$^{-+}$ spin-parity configuration. Although the
$\eta_{c}$ cannot be produced directly from e$^{+}$e$^{-}$
annihilations, it is produced copiously in radiative decays of
$J/\psi$ and $\psi^{\prime}$~\cite{PDG}. The large $J/\psi$ and
$\psi^{\prime}$ data samples taken with the BESIII detector at the
BEPCII provide an opportunity for a detailed study of $\eta_{c}$
decays.

The complexity of QCD remains unsolved in the charmonium-mass region,
and there are still many contradictions between pQCD calculations and
experimental measurements. In particular, the pQCD helicity selection
rule~\cite{SJB_GPL_PRD,VLC_ARZ_NPB,VLC_ARZ_PR} is violated in many
exclusive charmonium-decay processes, for example, the decay processes
with meson pairs in the final state, like $J/\psi\to VP$, $\eta_{c}\to
VV$, and $\chi_{c1}\to VV$, where $V$ and $P$ denote vector and
pseudoscalar mesons. Other examples include decay processes with baryon anti-baryon
pairs in the final state, such as $\eta_{c} \to B_{8}\bar{B}_{8}$, and
$\chi_{c0} \to B_{8}\bar{B}_{8}$, where $B_{8}$$\bar{B}_{8}$ denote
the octet baryon anti-baryon pairs. Many attempts have been made to
understand these contradictions, such as by the
quark-diquark model for the proton~\cite{MA_PRD_88, MA_PRD_91}, constituent quark-mass
corrections~\cite{MA_PRD_92, FM_PRD}, mixing between the charmonium state and the glueball~\cite{MA_PRD_94}, 
and the quark pair creation model~\cite{RGP_EPJA}.
 However, the measured branching fractions are not consistent with
 the predictions of any of these models.

In Refs.~\cite{YJZ_PRL,XHL_PRD}, intermediate meson loop (IML)
transitions are proposed, where the long-distance interaction can
evade the Okubo-Zweig-Iizuka (OZI) rule and allow the violation of the
pQCD helicity selection rule. Further calculations on the branching
fractions of $\eta_{c} \to B_{8}\bar{B}_{8}$, $\chi_{c0} \to
B_{8}\bar{B}_{8}$ and $h_{c} \to B_{8}\bar{B}_{8}$ based on
charmed-meson loops were carried out~\cite{XHL_JPG}, and the results
agree with the measured branching fractions of $\eta_{c} \to p\bar{p}$
and $\eta_{c} \to \Lambda\bar{\Lambda}$.  Using a sample of $2.25
\times 10^{8}$ $J/\psi$ events~\cite{JPSI_TTNM} collected with the
BESIII detector in 2009, we measure the branching fractions of $\eta_{c}
\to \Sigma^{+}\bar{\Sigma}^{-}$ and $\eta_{c} \to
\Xi^{-}\bar{\Xi}^{+}$ for the first time via the $J/\psi\to \gamma \eta_{c}$
radiative decay process.

\section{\boldmath{Detector and Monte Carlo simulation}}

BEPCII~\cite{BESIII} is a double-ring e$^{+}$e$^{-}$ collider designed to provide a
peak luminosity of $10^{33}$ cm$^{-2}s^{-1}$ at a center-of-mass energy of $3.77$~GeV.
 The BESIII~\cite{BESIII} detector has a geometrical acceptance of
$93\%$ of $4\pi$ and has four main components: (1) A small-cell,
helium-based ($40\%$ He, $60\%$ C$_{3}$H$_{8}$) main drift chamber
(MDC) with $43$ layers providing an average single-hit resolution of
$135$~$\mu$m, charged-particle momentum resolution in a $1$~T
magnetic field of $0.5\%$ at 1~GeV/$c$. (2) An electromagnetic calorimeter (EMC)
consisting of $6240$ CsI(Tl) crystals in a cylindrical structure
(barrel) and two endcaps. For $1$~GeV photons, the energy resolution is
$2.5\%$ ($5\%$) in the barrel (endcaps), and the position resolution
is $6$~mm ($9$~mm) in the barrel (endcaps). (3) A time-of-flight
system (TOF) consisting of $5$-cm-thick plastic scintillators, with
$176$ detectors of $2.4$~m length in two layers in the barrel and
$96$ fan-shaped detectors in the endcaps. The barrel (endcaps) time
resolution of $80$~ps ($110$~ps) provides $2\sigma$ $K/\pi$
separation for momenta up to $\sim 1$~GeV/$c$. (4) The muon system
(MUC) consists of $1000$~m$^{2}$ of Resistive Plate Chambers (RPCs)
in $9$ barrel and $8$ endcap layers and provides $2$~cm position
resolution.

The optimization of the event selection and the estimate of backgrounds
 are performed using Monte Carlo (MC) simulated data. The {\sc GEANT4}~\cite{GEANT}-based
simulation software {\sc BOOST}~\cite{BOOST} includes the geometry and the material description
of the BESIII spectrometer, the detector response and digitization models, as
well as the tracking of the detector running conditions and performances.
The production of the $J/\psi$ resonance is simulated by the MC event generator
{\sc KKMC}~\cite{KKMC1, KKMC2}, while the decays are generated by {\sc EvtGen}~\cite{EvtGen} for
the known decay modes with branching fractions set to world average values~\cite{PDG},
and by {\sc LundCharm}~\cite{LUNDCHARM} for the remaining unknown decays.

\section{\boldmath{Event selection}}

We select $\eta_{c}$ mesons via the radiative decay $J/\psi \to \gamma
\eta_{c}$ with its subsequent decay into $\Sigma^{+}\bar{\Sigma}^{-}$
and $\Xi^{-}\bar{\Xi}^{+}$. The $\Sigma^{+}$ candidates are
reconstructed from the decay $\Sigma^{+}\to p\pi^{0}$ with the
$\pi^{0}$ decaying into a pair of photons; the $\Xi^{-}$ candidates
are reconstructed from the decays $\Xi^{-} \to \Lambda\pi^{-}$ and
$\Lambda\to p\pi^{-}$. The anti-particle candidates,
$\bar{\Sigma}^{-}$ and $\bar{\Xi}^{+}$, are reconstructed in a similar
way but with the decay products changed to the corresponding
anti-particles.

\begin{figure}[h!]
\begin{center}
\includegraphics[scale=0.31]{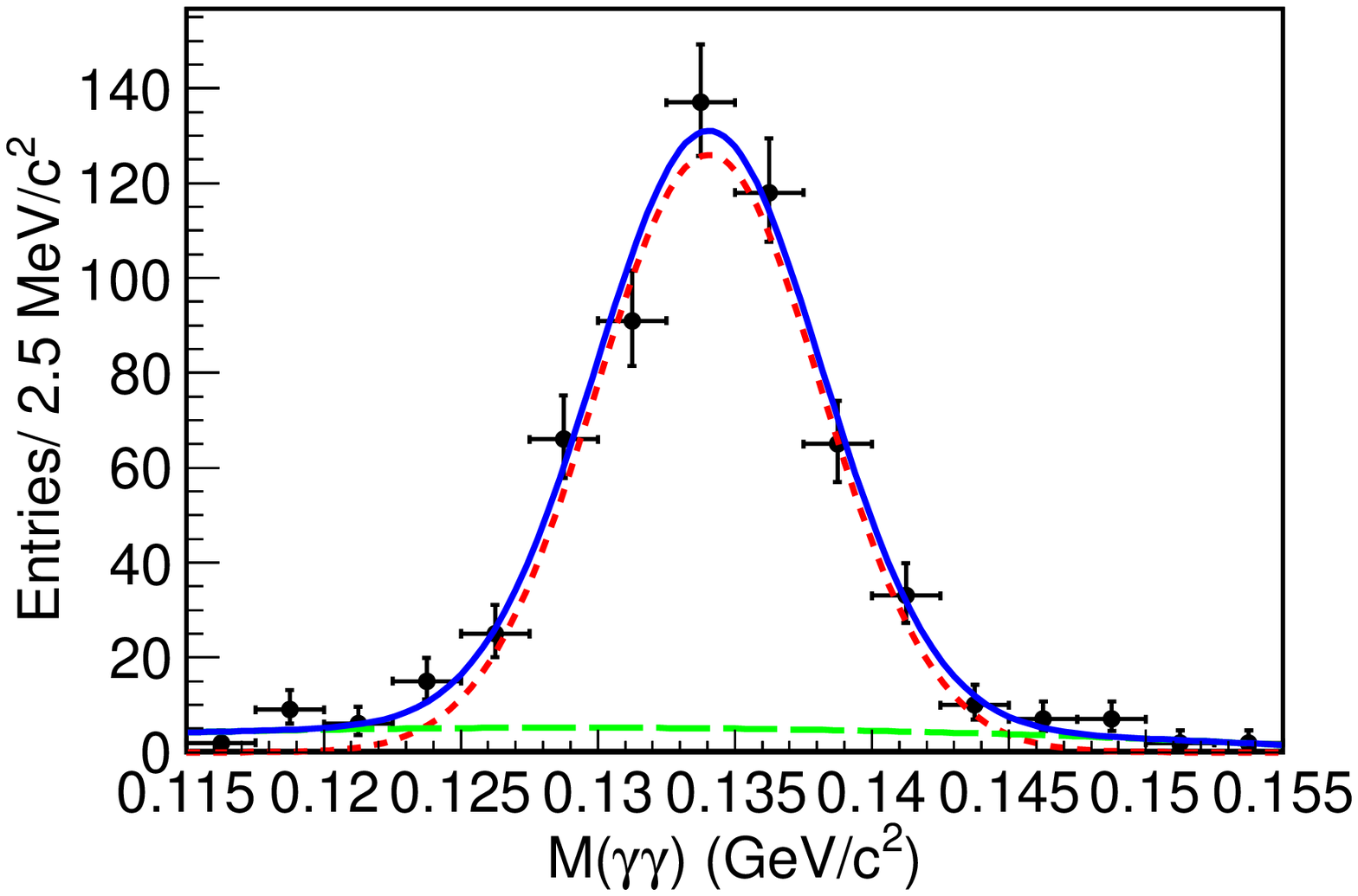}\put(-40,82){(a)}\includegraphics[scale=0.31]{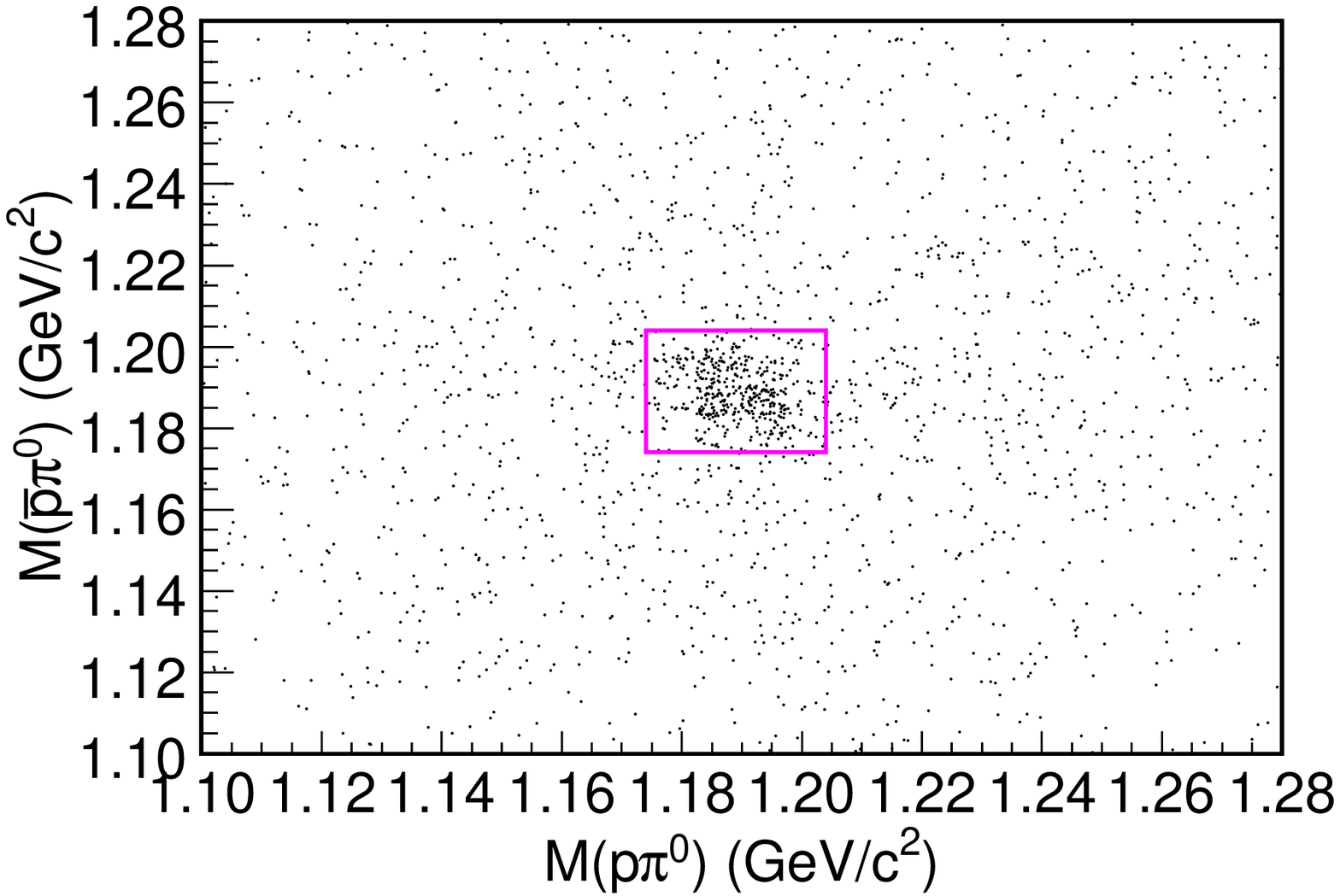}\put(-40,82){(b)} \includegraphics[scale=0.31]{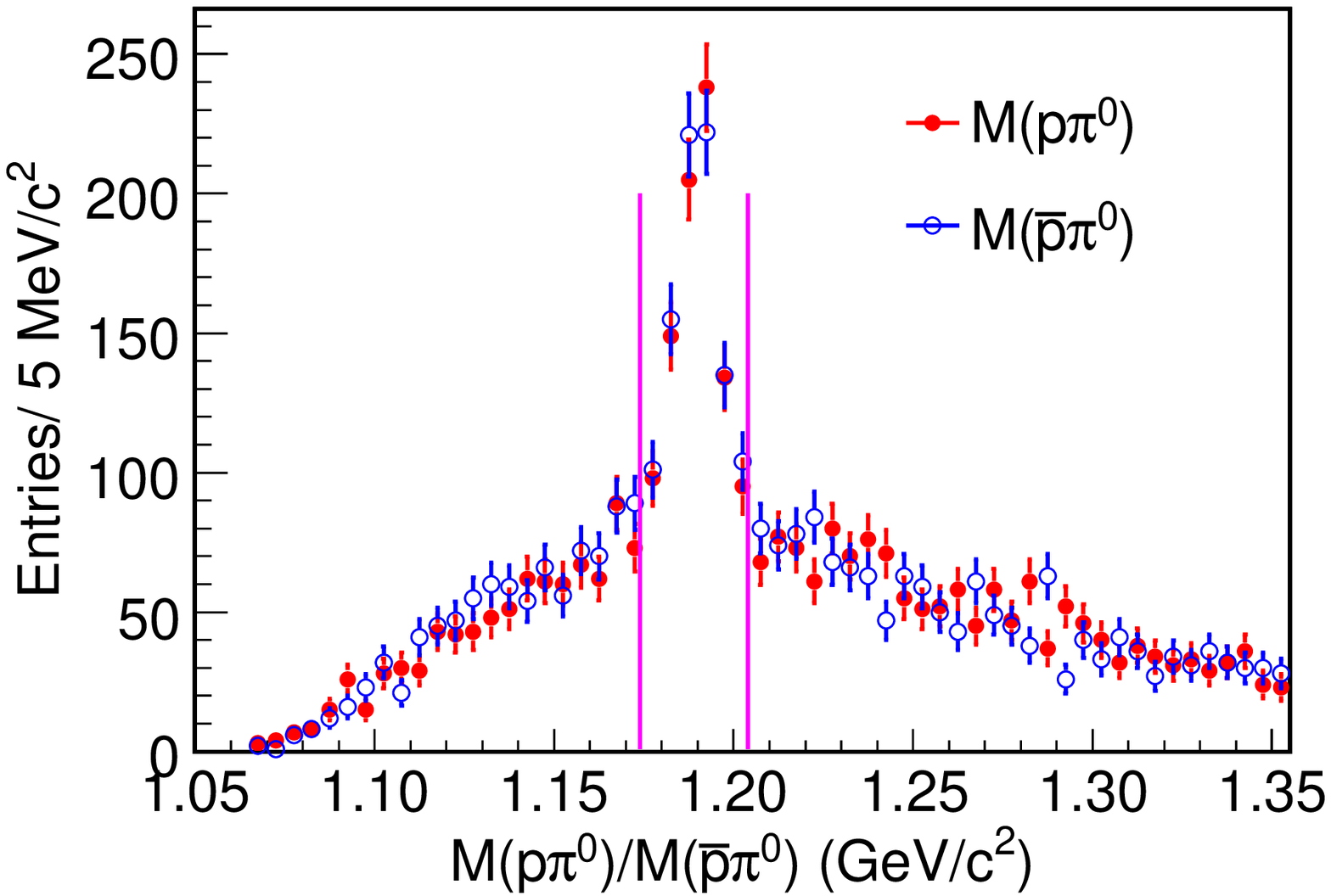}\put(-40,82){(c)}
\caption{(a) A fit to the invariant-mass distribution of
  $\gamma\gamma$ pairs after applying all the requirements except for
  the $\pi^{0}$-mass window. Dots with error bars are data, and the
  solid line is the total fit result. The signal is represented by the
  short-dashed line and the background by the long-dashed line. (b) A
  scatter plot for M$(\bar{p}\pi^{0})$ versus M$(p\pi^{0})$. (c)
  Invariant-mass distributions of $p\pi^{0}$ and $\bar{p}\pi^{0}$;
  solid dots with error bars are M$(p\pi^{0})$, and the open circles with
  error bars are M$(\bar{p}\pi^{0})$.}
\label{fig:sigmaresonances}
\end{center}
\end{figure}

Tracks of charged particles in the polar-angle range
$|\cos\theta|<0.93$ are reconstructed from hits in the MDC. The TOF
and $dE/dx$ information are combined to form particle identification
(PID) confidence levels for the $\pi$,~$K$ and $p$ hypotheses. Each
track is assigned to the particle type that corresponds to the
hypothesis with the highest confidence level. Photon candidates are
reconstructed by clustering the energy deposited in the EMC
crystals. The minimum energy requirement is $25$~MeV for barrel
showers ($|\cos\theta|<0.80$) and $50$~MeV for endcap showers
($0.86<|\cos\theta|<0.92$). Requirements on the EMC cluster timing
are applied to suppress electronic noise and energy deposits unrelated
to the event. Candidate $\pi^{0}$ mesons are reconstructed from pairs
of photons with an invariant mass in the range $0.115$~GeV/$c^{2}<$
M$(\gamma\gamma) <0.155$~GeV/$c^{2}$. The $\pi^{0}$ invariant-mass resolution
is determined to be $4.2$ MeV/$c^{2}$ by fitting the invariant-mass
distribution of the $\gamma\gamma$ pairs from data after applying all
the requirements except for the $\pi^{0}$-mass window, as shown in
Fig.~\ref{fig:sigmaresonances}(a). In the fit, the $\pi^{0}$ signal
is taken with a Gaussian form, and the background is described by a
second-order Chebychev polynomial function.

\begin{figure}[h!]
\begin{center}
\includegraphics[scale=0.31]{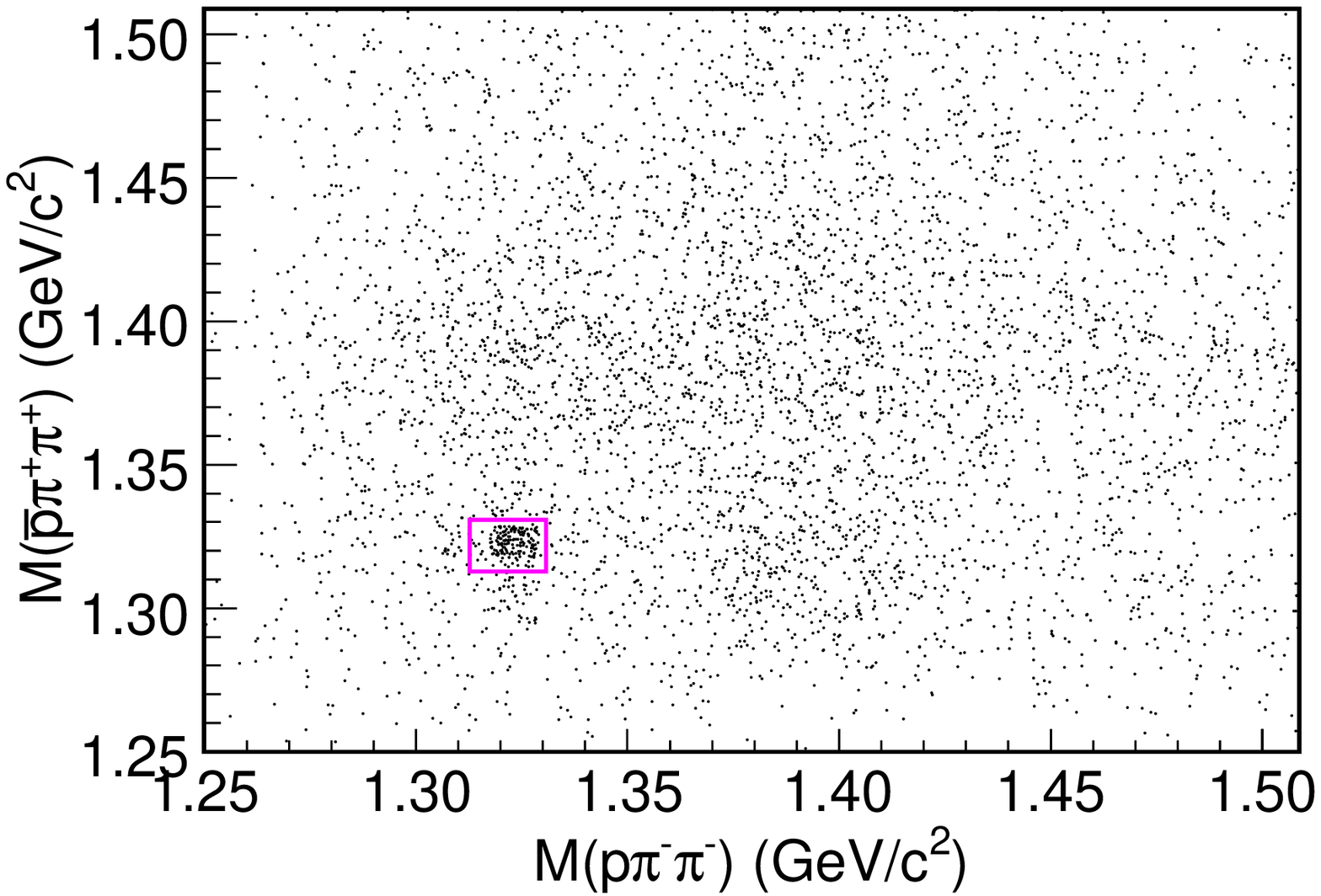}\put(-40,80){(a)} \includegraphics[scale=0.31]{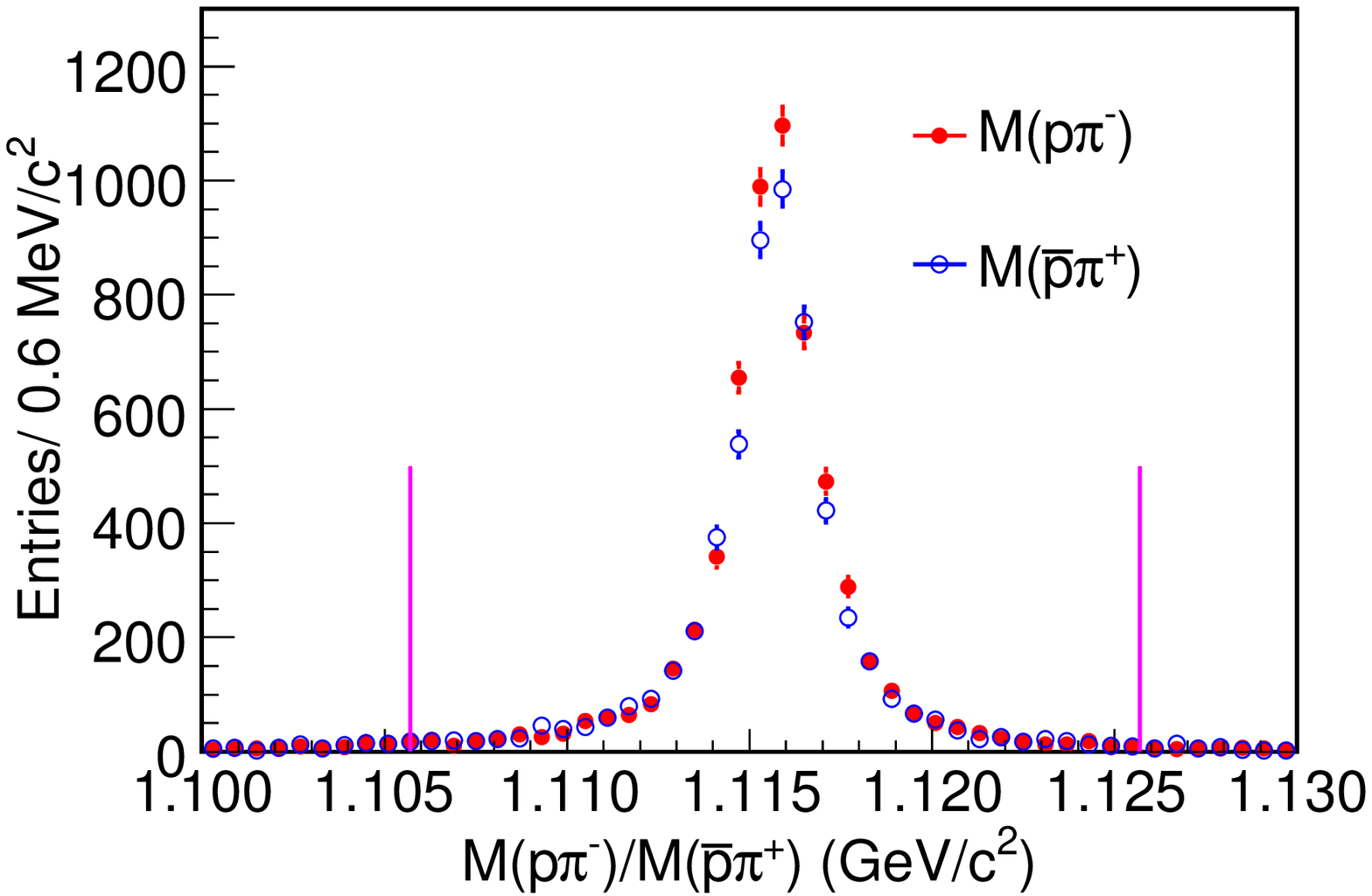}\put(-40,80){(b)}\includegraphics[scale=0.31]{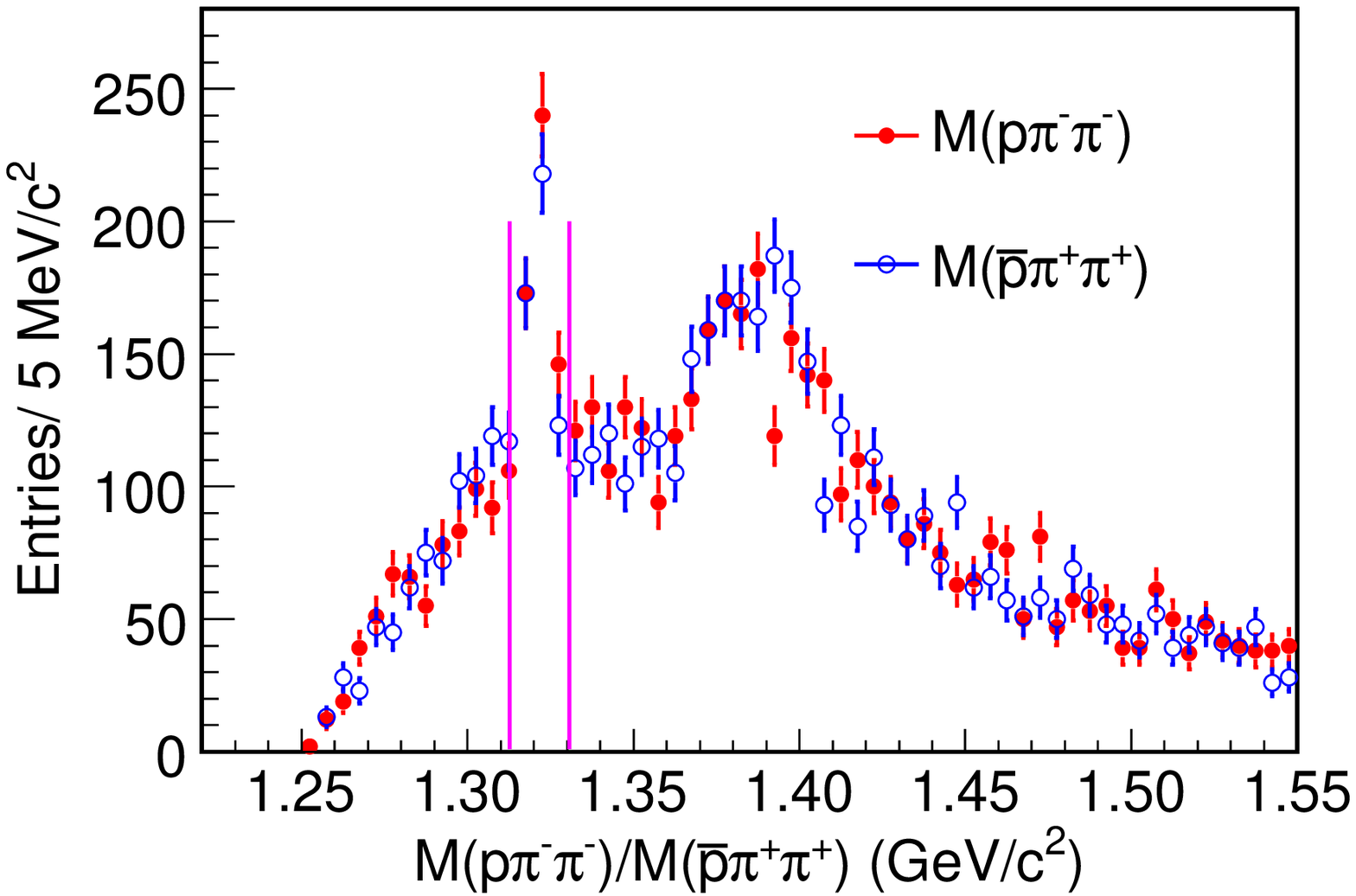}\put(-40,80){(c)}
\caption{(a) Scatter plot for M$(\bar{p}\pi^{+}\pi^{+})$ versus
  M$(p\pi^{-}\pi^{-})$. Invariant-mass distributions of (b) $p\pi^{-}$
  and $\bar{p}\pi^{+}$, and (c) $p\pi^{-}\pi^{-}$ and
  $\bar{p}\pi^{+}\pi^{+}$. Solid dots with error bars are
  M$(p\pi^{-})$ and M$(p\pi^{-}\pi^{-})$, and open circles with error
  bars are M$(\bar{p}\pi^{+})$ and M$(\bar{p}\pi^{+}\pi^{+})$. }
\label{fig:xiresonances}
\end{center}
\end{figure}

For $J/\psi \to \gamma\eta_{c} \to \gamma\Sigma^{+}\bar{\Sigma}^{-}$,
exactly one proton, one anti-proton, at least five photons and at
least two $\pi^{0}$ candidates from the combination of these photons
are required.  A four-constraint (4C) kinematic fit, based on momentum
and energy conservation, is applied under the $J/\psi \to \gamma
p\bar{p}\pi^{0}\pi^{0}$ hypothesis, and $\chi^{2}_{4C}<30$ is
required. For events with more than five photons or more than two
$\pi^{0}$ candidates, the combination with the minimum $\chi^{2}_{4C}$
is retained in the analysis. The events are also fitted to the $J/\psi
\to p\bar{p}\pi^{0}\pi^{0}$ and $J/\psi \to \gamma\gamma
p\bar{p}\pi^{0}\pi^{0}$ hypotheses.  We require
$\chi^{2}_{4C}(p\bar{p}\pi^{0}\pi^{0}) > 200$ and
$\chi^{2}_{4C}(\gamma p\bar{p}\pi^{0}\pi^{0}) <
\chi^{2}_{4C}(\gamma\gamma p\bar{p}\pi^{0}\pi^{0})$.  The
$p$,~$\bar{p}$ and the two $\pi^{0}$ candidates are combined to form
the $\Sigma^{+}$ and $\bar{\Sigma}^{-}$ candidates by minimizing
$($M$_{p\pi^{0}_{1}}-$M$_{\Sigma^{+}})^{2}+($M$_{\bar{p}\pi^{0}_{2}}
-$M$_{\bar{\Sigma}^{-}})^{2}$. Furthermore, the combined
$p$,~$\pi^{0}$ ($\bar{p}$,~$\pi^{0}$) pair must have an invariant mass
within $15$~MeV/$c^{2}$ of the $\Sigma^{+}$ ($\bar{\Sigma}^{-}$) mass,
as shown in Fig.~\ref{fig:sigmaresonances}(b) and (c).

For $J/\psi \to \gamma\eta_{c} \to \gamma\Xi^{-}\bar{\Xi}^{+}$,
exactly one proton, one anti-proton, two $\pi^{+}$s, two $\pi^{-}$s
and at least one photon are required. A 4C kinematic fit is applied
under the $J/\psi \to \gamma p\bar{p}\pi^{+}\pi^{+}\pi^{-}\pi^{-}$
hypothesis, and $\chi^{2}_{4C} < 90$ is required. For events with more
than one photon candidate, only the combination with the minimum
$\chi^{2}_{4C}$ is retained in the analysis.  The events are also
fitted to the $J/\psi \to p\bar{p}\pi^{+}\pi^{+}\pi^{-}\pi^{-}$ and
$J/\psi \to \gamma\gamma p\bar{p}\pi^{+}\pi^{+}\pi^{-}\pi^{-}$
hypotheses.  We require
$\chi^{2}_{4C}(p\bar{p}\pi^{+}\pi^{+}\pi^{-}\pi^{-})>200$ and
$\chi^{2}_{4C}(\gamma p\bar{p}\pi^{+}\pi^{+}\pi^{-}\pi^{-})<
\chi^{2}_{4C}(\gamma \gamma p\bar{p}\pi^{+}\pi^{+}\pi^{-}\pi^{-})$.

To reconstruct the kinematical information of $\Lambda$ and $\Xi^{-}$,
vertex fits are applied to the charged tracks ($p\pi^{-}$ and
$p\pi^{-}\pi^{-}$ for $\Lambda$ and $\Xi^{-}$, respectively), with the
requirement that all the tracks originated from the same decay point.
Next, secondary vertex fits are applied to these reconstructed
particles, with the requirement that their flight time is consistent
with the one predicted from their final-state particles. The
$p$,~$\pi^{-}$ ($\bar{p}$,~$\pi^{+}$) combination with an invariant
mass that is the closest to the $\Lambda$ ($\bar{\Lambda}$) mass is
chosen to form the $\Lambda$ ($\bar{\Lambda}$). Furthermore, the mass
difference must be within $10$~MeV/$c^{2}$, as shown in
Fig.~\ref{fig:xiresonances}(b). The $p$, $\pi^{-}$, $\pi^{-}$
($\bar{p}$, $\pi^{+}$, $\pi^{+}$) combination must have an invariant
mass within $9$ MeV/$c^{2}$ of the $\Xi^{-}$ ($\bar{\Xi}^{+}$)
mass, as shown in Fig.~\ref{fig:xiresonances}(a) and (c).

Figure~\ref{fig:fits} shows the invariant-mass distributions of
$\Sigma^{+}\bar{\Sigma}^{-}$ and $\Xi^{-}\bar{\Xi}^{+}$ pairs after
applying all the event selection criteria.  A clear signature of an
$\eta_{c}$ resonance is observed.

\section{\boldmath{Background studies}}

The background can be classified into two categories: background from
$\eta_{c}$ decays which produces a peak within the $\eta_{c}$ signal region,
and background from $J/\psi$ decays which gives a smooth distribution under
the $\eta_{c}$ resonance.

\begin{figure}[h!]
\begin{center}
\includegraphics[scale=0.45]{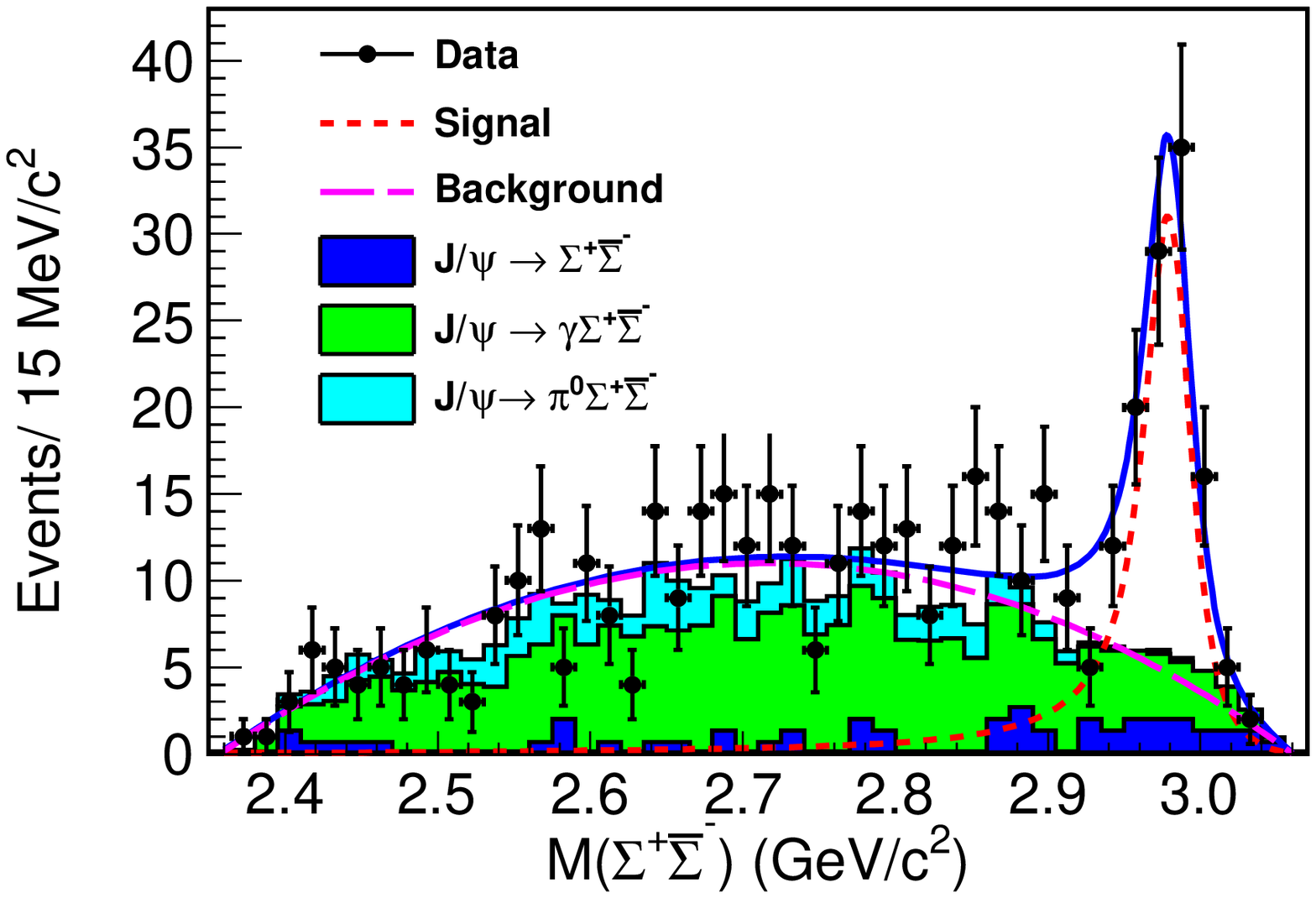}\put(-52,130){(a)} \includegraphics[scale=0.45]{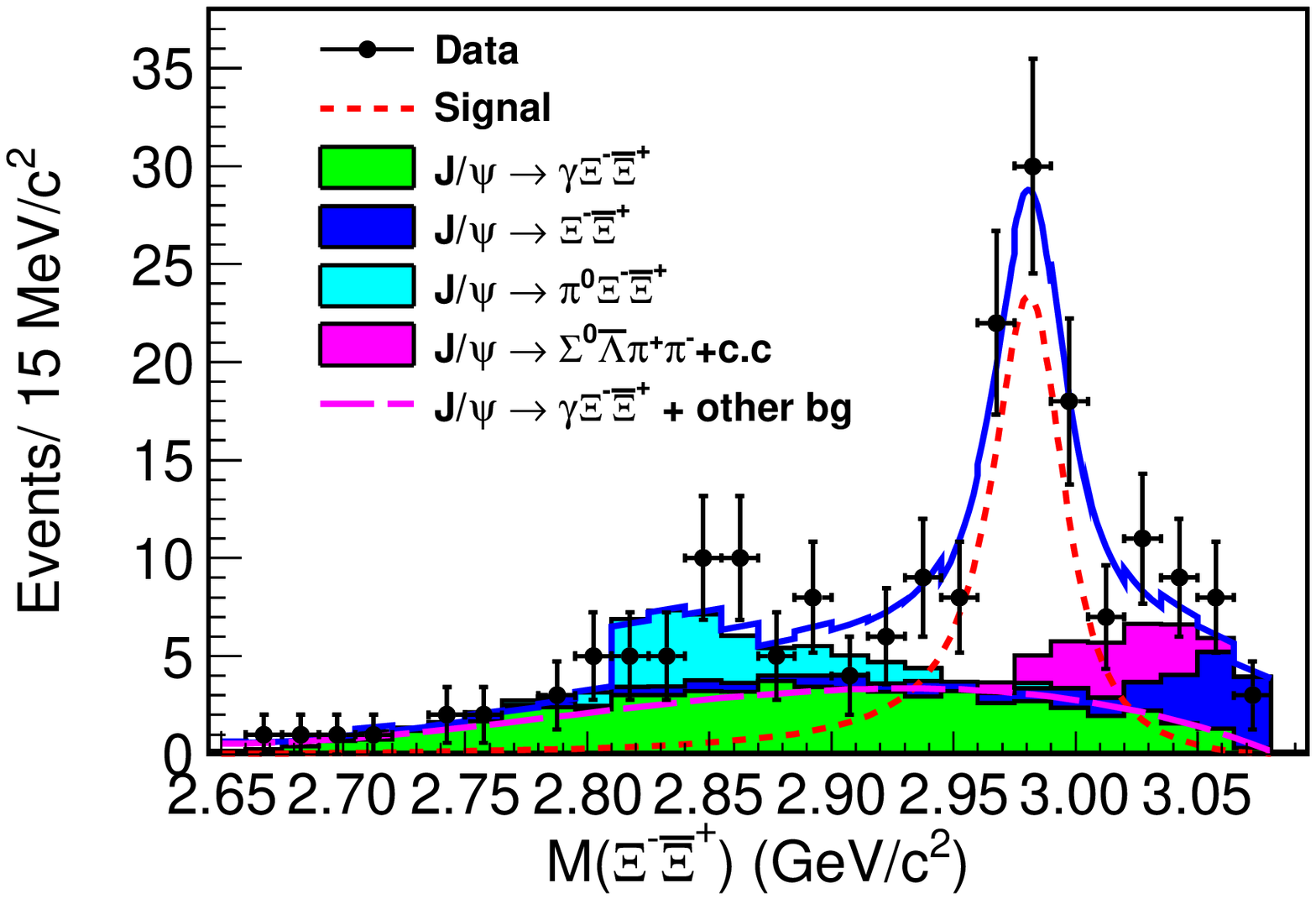}\put(-52,130){(b)}
\caption{Invariant mass distributions of data and MC background
  channels together with the fitted curves for (a)
  $\Sigma^{+}\bar{\Sigma}^{-}$, and (b) $\Xi^{-}\bar{\Xi}^{+}$. Dots
  with error bars are data, and the histograms are the backgrounds
  from simulated $J/\psi$ decays.
  Solid lines are the total fit results, signals
  are shown in short-dashed lines, and backgrounds are shown as
  long-dashed lines and shaded histograms.}

\label{fig:fits}
\end{center}
\end{figure}

For $\eta_{c} \to \Sigma^{+}\bar{\Sigma}^{-}$, the
potential peaking background channel is $\eta_{c} \to
p\bar{p}\pi^{0}\pi^{0}$, which has not previously been measured.
By requiring the invariant mass of any $p\pi^{0}$
combination to be outside a mass window of $50$ MeV/$c^{2}$ centered
at the $\Sigma^{+}$ mass and the $p\bar{p}\pi^{0}\pi^{0}$ invariant
mass within $30$ MeV/$c^{2}$ from the $\eta_{c}$ mass, the number of
$\eta_{c} \to p\bar{p}\pi^{0}\pi^{0}$ events is obtained, and the
branching fraction is determined to be $(5.0\pm0.6_{\rm stat.})\times10^{-4}$, where the uncertainty is statistical only.
Out of $5\times10^{5}$ $J/\psi \to \gamma\eta_{c} \to \gamma p\bar{p}\pi^{0}\pi^{0}$ MC simulated events, $193$ events survive
after applying the event selection criteria.
Using the measured branching fraction, the background contribution from this process is estimated to be $0.7$ events. For the background from $J/\psi$ decays, the main sources are
$J/\psi \to \Sigma^{+}\bar{\Sigma}^{-}$ and $J/\psi \to
\pi^{0}\Sigma^{+}\bar{\Sigma}^{-}$, which have a fake photon or a
photon from $\pi^{0}$ that escaped from detection, respectively; and
$J/\psi \to \gamma \Sigma^{+}\bar{\Sigma}^{-}$, which is an
irreducible background to the signal process. Using $5\times10^{5}$ MC
simulated events for each channel and applying the event selection
criteria to these MC samples, the background contributions are
estimated by normalizing the number of the surviving events to the
total number of $J/\psi$ events. In the normalization, the branching
fraction of $J/\psi \to \Sigma^{+} \bar{\Sigma}^{-}$ is taken from
Ref.~\cite{PDG}, and the branching fractions of
$J/\psi \to \gamma \Sigma^{+}\bar{\Sigma}^{-}$ and $J/\psi \to \pi^{0}
\Sigma^{+}\bar{\Sigma}^{-}$ are measured in this analysis. The
branching fraction of $J/\psi \to \pi^{0} \Sigma^{+}\bar{\Sigma}^{-}$
is measured to be $(5.0\pm0.1_{\rm stat.})\times10^{-4}$ using similar
event selection criteria but with an additional photon and a
$\pi^{0}$ reconstructed from the selected photons.
The branching fraction of $J/\psi \to
\gamma\Sigma^{+}\bar{\Sigma}^{-}$ is measured to be $(7.4\pm0.6_{\rm
  stat.})\times10^{-5}$ with the same event selection criteria as was
applied for the signal events, but without requiring that the
$\Sigma^{+}\bar{\Sigma}^{-}$ system forms an $\eta_{c}$ resonance and
with a selection on the invariant mass of $2.4$ GeV/$c^{2}$ $<$
M$(\Sigma^{+}\bar{\Sigma}^{-})$ $<$ 2.8 GeV/$c^{2}$.  The total
background is estimated to be $351$ events in the entire mass region,
as shown in Fig.~\ref{fig:fits}(a). The total background shape is
found to be smooth without an enhancement under the $\eta_{c}$
resonance.

For $\eta_{c} \to \Xi^{-}\bar{\Xi}^{+}$, the potential peaking
background channels are $\eta_{c} \to
p\bar{p}\pi^{+}\pi^{+}\pi^{-}\pi^{-}$ and $\eta_{c} \to
\Lambda\bar{\Lambda}\pi^{+}\pi^{-}$. Out of $2.5\times10^{5}$
simulated MC events for each channel, $2$ and $21$ events survived
after applying the event selection criteria.
The branching fractions of
these two channels are determined to be $(6.7\pm1.0_{\rm stat.})\times10^{-4}$ and $(6.3\pm0.4_{\rm stat.})\times10^{-3}$, respectively, where the uncertainties are statistical only. The invariant-mass requirements for $\eta_{c}\to p\bar{p}\pi^{+}\pi^{+}\pi^{-}\pi^{-}$ are:
$|$M$_{p\pi^{-}}-$M$_{\Lambda}|>20$ MeV/$c^{2}$ (no $p\pi^{-}$
combination consistent with a $\Lambda$),
$|$M$_{p\pi^{-}\pi^{-}}-$M$_{\Xi^-}|>25$ MeV/$c^{2}$ (no
$p\pi^{-}\pi^{-}$ combination consistent with a $\Xi^-$), and
$|$M$_{p\bar{p}\pi^{+}\pi^{+}\pi^{-}\pi^{-}}-$M$_{\eta_{c}}|<30$
MeV/$c^{2}$; for $\eta_{c} \to \Lambda\bar{\Lambda}\pi^{+}\pi^{-}$,
the only change is $|$M$_{p\pi^{-}}-$M$_{\Lambda}|<20$
MeV/$c^{2}$. Using the measured branching fractions, the background
contributions from the two peaking background channels are estimated
to be $0.02$ and $2$ events to the signal after normalizing the number
of the surviving events to the total number of the $J/\psi$ events,
respectively. The main background channels from $J/\psi$ decays are
$J/\psi \to \Xi^{-}\bar{\Xi}^{+}$ and $J/\psi \to
\pi^{0}\Xi^{-}\bar{\Xi}^{+}$, which have one fake photon or one photon
from the $\pi^{0}$ that escaped from detection, and $J/\psi \to \gamma
\Xi^{-}\bar{\Xi}^{+}$, which is an irreducible background to the
signal. Another background contribution from $J/\psi \to
\Sigma^{0}\bar{\Lambda} \pi^{+}\pi^{-}\to \gamma
\Lambda\bar{\Lambda}\pi^{+}\pi^{-}+c.c.$ is apparently seen from the
invariant-mass distribution of $\gamma\Lambda$ pairs. To estimate the
background contribution from the process $J/\psi \to
\pi^{0}\Xi^{-}\bar{\Xi}^{+}$ including intermediate states, $J/\psi
\to \pi^{0}\Xi^{-}\bar{\Xi}^{+}$ decays are reconstructed from data,
and the signal yield is obtained in each M($\Xi^{-}\bar{\Xi}^{+}$)
mass bin.  The selection criteria are similar to that for signal
events but with an additional photon and a $\pi^{0}$ reconstructed from the selected photons.
 The relative efficiencies of the $\gamma \Xi^{-}\bar{\Xi}^{+}$ and
$\pi^{0}\Xi^{-}\bar{\Xi}^{+}$ selection criteria are estimated in each
M($\Xi^{-} \bar{\Xi}^{+}$) mass bin using $J/\psi \to \pi^{0}\Xi^{-}
\bar{\Xi}^{+}$ MC events. Combining this relative efficiency with the
number of $J/\psi \to \pi^{0}\Xi^{-} \bar{\Xi}^{+}$ signal events in
each M($\Xi^{-}\bar{\Xi}^{+}$) mass bin, the number of $\pi^{0}\Xi^{-}
\bar{\Xi}^{+}$ events that pass the $\gamma\Xi^{-}\bar{\Xi}^{+}$
selection is estimated. We generated $5\times10^{6}$ MC events for the
channels $J/\psi \to \Xi^{-}\bar{\Xi}^{+}$ and $J/\psi \to
\Sigma^{0}\bar{\Lambda} \pi^{+}\pi^{-}+c.c.$ and $2.5\times10^{5}$ MC
events for the channel of $J/\psi \to \gamma \Xi^{-}\bar{\Xi}^{+}$,
and applied the event selection criteria to these MC samples. The
contribution from each background process is estimated by normalizing
the number of the surviving events to the total number of the $J/\psi$
events.  In the normalization, the branching fraction of $J/\psi \to
\Xi^{-}\bar{\Xi}^{+}$ is taken from Ref.~\cite{PDG} and the branching
fractions of $J/\psi \to \gamma \Xi^{-}\bar{\Xi}^{+}$ and $J/\psi \to
\Sigma^{0}\bar{\Lambda}\pi^{+}\pi^{-}$ are measured in this analysis.
The branching fraction of $J/\psi \to
\Sigma^{0}\bar{\Lambda}\pi^{+}\pi^{-}$ is measured to be
$(4.7\pm0.1_{\rm stat.})\times10^{-4}$ by fitting the invariant-mass
distribution of $\gamma\Lambda$ pairs. The branching fraction of
$J/\psi \to \gamma \Xi^{-} \bar{\Xi}^{+}$ is measured to be $(1.8\pm
0.5_{\rm stat.})\times10^{-5}$ by excluding the $\Xi^{-}\bar{\Xi}^{+}$
system to form an $\eta_{c}$ meson via the requirement
M($\Xi^{-}\bar{\Xi}^{+})<2.8$ GeV/$c^{2}$. The total background from
$J/\psi$ decays is estimated to be $116$ events in the entire mass
region, as shown in Fig.~\ref{fig:fits}(b), and is smoothly
distributed and no enhancement under the $\eta_{c}$ resonance is
observed.

\section{\boldmath{SIGNAL EXTRACTIONS AND BRANCHING FRACTION CALCULATIONS}}

Signal yields are obtained from unbinned maximum likelihood fits to
the invariant-mass distributions of $\Sigma^{+}\bar{\Sigma}^{-}$ and
$\Xi^{-}\bar{ \Xi}^{+}$ candidates.
The probability density function (PDF) used
in the fit is given by
\begin{equation*}
F(m)=\sigma_{res}\otimes (\varepsilon(m)\times E_{\gamma}^{3} \times damping(E_{\gamma}) \times BW(m)) + BKG(m),
\end{equation*}
where $BW(m)$ and $BKG(m)$ are the signal component described by the
Breit-Wigner form and the background component, respectively;
$\sigma_{res}$ is the experimental resolution function and
$\varepsilon(m)$ is the mass-dependent efficiency; $E_{\gamma}^{3}$ is
the cube of the radiative photon energy and reflects the expected
energy dependence of the magnetic-dipole (M1) matrix element;
$damping(E_{\gamma})$ describes a function to damp the diverging tail
caused by the $E_{\gamma}^{3}$ dependence and is given in the form of
$\frac{E_{0}^{2}}{E_{\gamma}E_{0}+(E_{\gamma}-E_{0})^{2}}$ as used by
KEDR~\cite{KEDR}, where $E_{0}$ is the peak energy of the transition
photon.

The experimental resolution function is determined from a signal MC
sample with the width of the $\eta_{c}$ set to zero. A double
Gaussian function is used for $\eta_{c} \to
\Sigma^{+}\bar{\Sigma}^{-}$ and a single Gaussian function for
$\eta_{c}\to \Xi^{-}\bar{\Xi}^{+}$. The mass-dependent efficiencies
are determined from phase-space MC samples. The background component
in the channel $\eta_{c} \to \Sigma^{+}\bar{\Sigma}^{-}$ is described
by a third-order polynomial function. The background in the channel
$\eta_{c} \to \Xi^{-}\bar{\Xi}^{+}$ is composed of four parts: (1)
contributions of $J/\psi \to \Xi^{-}\bar{\Xi}^{+}$, $J/\psi\to
\pi^{0}\Xi^{-}\bar{\Xi}^{+}$ and $J/\psi \to \Sigma^{0}\bar{
  \Lambda}\pi^{+}\pi^{-}+c.c.$, with shapes and normalizations fixed in
the fit; (2) a third-order Chebychev polynomial function representing
the phase-space background contribution from $J/\psi \to
\gamma\Xi^{-}\bar{\Xi}^{+}$ and other possible processes, with
parameters set free in the fit.

The signal detection efficiency is determined with MC simulated events
by comparing the number of events after the event selection with the
number of generated events.  In the simulation, the decay
$J/\psi \to \gamma\eta_{c}$ is generated using the 
helicity amplitude method~\cite{HELAMP},
and the radiative photon follows the angular distribution of
$1+\cos^{2}(\theta)$, where $\theta$ is the polar angle of the
radiative photon.  The final state  
baryons' angular distributions are assumed to be uniformly
distributed in the rest frame of the $\eta_{c}$.

The fitted curves are shown in Fig.~\ref{fig:fits} for $\eta_{c} \to
\Sigma^{+}\bar{\Sigma}^{-}$ and $\eta_{c} \to \Xi^{-}\bar{\Xi}^{+}$,
where the mass and width of the $\eta_{c}$ are fixed to the
newly measured results from BESIII~\cite{BESIII_ETAC}. A possible
interference between the $\eta_{c}$ resonance amplitude and the
non-resonant background is neglected.  The observed number of events,
$N_{obs}$, are listed in Table~\ref{tab:result}. The
statistical significances of the signals are calculated
using the changes in the log-likelihood values and the number of degrees of
freedom of the fits with and without the $\eta_{c}$ signal
assumptions. For $\eta_c \to \Sigma^+\bar{\Sigma}^-$, the change in $-ln(\mathcal{L})$ with $\Delta(d.o.f.) = 1$ is $43.2$, corresponding to a statistical significance of $9.3\sigma$. For $\eta_c \to \Xi^-\bar{\Xi}^+$, the change in $-ln(\mathcal{L})$ with $\Delta(d.o.f.) = 1$ is $20.2$, corresponding to a statistical significance of $6.4\sigma$. The branching fraction of $\eta_{c} \to
\Sigma^{+}\bar{\Sigma}^{-}$ is calculated with:
\begin{eqnarray*}
\mathcal{B}(\eta_{c} \to \Sigma^{+}\bar{\Sigma}^{-}) = \frac{N_{obs}-N_{peaking}}{N_{J/\psi}\times \mathcal{B}(J/\psi \to \gamma\eta_{c})\times \mathcal{B}^{2}(\Sigma^{+} \to p\pi^{0})\times \mathcal{B}^{2}(\pi^{0} \to \gamma\gamma)\times\varepsilon},
\end{eqnarray*}
where $N_{peaking}$ is the number of peaking background events
determined from the background study, $N_{J/\psi}$ is the total number
of $J/\psi$ events, which is $2.25\times10^{8}$ with an uncertainty of
$1.2\%$~\cite{JPSI_TTNM}, $\mathcal{B}(J/\psi \to \gamma\eta_{c})$,
$\mathcal{B}(\Sigma^{+} \to p\pi^{0})$ and $\mathcal{B}(\pi^{0} \to
\gamma\gamma)$ are the branching fractions of $J/\psi \to \gamma
\eta_{c}$, $\Sigma^{+}\to p\pi^{0}$ and $\pi^{0} \to \gamma\gamma$,
respectively~\cite{PDG}, and $\varepsilon$ is the total detection efficiency. The
branching fraction of $\eta_{c} \to \Xi^{-}\bar{\Xi}^{+}$
is calculated with:
\begin{eqnarray*}
\mathcal{B}(\eta_{c} \to \Xi^{-}\bar{\Xi}^{+}) = \frac{N_{obs}-N_{peaking}}{N_{J/\psi}\times \mathcal{B}(J/\psi \to \gamma\eta_{c})\times \mathcal{B}^{2}(\Xi^{-} \to \Lambda\pi^{-})\times \mathcal{B}^{2}(\Lambda \to p\pi^{-})\times\varepsilon},
\end{eqnarray*}
where $\mathcal{B}(\Xi^{-} \to \Lambda\pi^{-})$ and $\mathcal{B}(\Lambda \to p\pi^{-})$ are the
branching fractions of $\Xi^{-} \to \Lambda\pi^{-}$ and $\Lambda \to p\pi^{-}$, respectively~\cite{PDG}.
The results are summarized in Table~\ref{tab:result}.

\begin{table}[h!]
\begin{center}
  \caption{Branching fractions of $\eta_{c} \to
    \Sigma^{+}\bar{\Sigma}^{-}$ and $\eta_{c} \to
    \Xi^{-}\bar{\Xi}^{+}$ obtained from this analysis and the
    predictions based on IML. For the measured branching fractions,
    the first uncertainty is statistical, the second experimental systematic,
    and the third is from input branching fractions taken from Ref.~\cite{PDG}.
}
\begin{tabular}{l|c|c}
\hline \hline
            &   $\eta_{c} \to \Sigma^{+}\bar{\Sigma}^{-}$   &   $\eta_{c} \to \Xi^{-}\bar{\Xi}^{+}$  \\
\hline

Statistical significance       &  $9.3\sigma$                    &               $6.4\sigma$                  \\

$N_{obs}$   &                    $112\pm15$                      &               $78\pm14$                    \\

$N_{peaking}$  &                    $0.7$                        &                 $2.0$                      \\

$\varepsilon$      &               $5.3\%$                       &                 $5.5\%$                    \\

Branching fraction ($10^{-3}$) &   $2.11\pm0.28\pm0.18\pm0.50$           &     $0.89\pm0.16\pm0.08\pm0.21$  \\

Branching fraction based on IML~\cite{XHL_JPG} ($10^{-3}$) & $0.51-1.00$    &             $0.48-0.96$          \\
\hline \hline
\end{tabular}
\label{tab:result}
\end{center}
\end{table}

\section{\boldmath{Systematic uncertainties}}

The sources of systematic uncertainties for the two measurements
are mainly from errors in the branching fractions of the known
intermediate decay modes; the reconstruction and identification
efficiencies of charged particles; the photon
reconstruction; the $\pi^{0}$, $\Sigma^{+}$, $\Lambda$
and $\Xi^{-}$ selection; vertex fits and kinematic fits; the fitting to
the invariant-mass distributions; event generators and the total
number of the $J/\psi$ events. The contributions are summarized in
Table~\ref{tab:SE_SIGMA_XI}.

The tracking and identification efficiency of protons from the
$\Sigma^{+}$ decay is determined using the $J/\psi \to
\Sigma^{+}\bar{\Lambda}\pi^{-}$ data sample. The recoiling mass
distribution of $\bar{\Lambda}\pi^{-}$ pairs is fitted to obtain the
$\Sigma^{+}$ signal yield, and the ratio between the yields with and
without the requirement of tracking and identifying the proton from
the $\Sigma^{+}$ decay is determined. The tracking and PID efficiency
for simulated MC events agrees within $2.0\%$ with that obtained from
the experimental data for each charged track. Hence, adding the
uncertainties of the proton and anti-proton in quadrature, $2.8\%$ is
taken as the systematic uncertainty from reconstructing the final
state charged tracks and their identification for
$\eta_{c} \to \Sigma^{+}\bar{\Sigma}^{-}$.

The tracking and PID efficiencies of $p$, $\bar{p}$, $\pi^{+}$ and
$\pi^{-}$ from $\Xi^{-}$ and $\bar{\Xi}^{+}$ decays are determined
from analyzing $J/\psi \to \Xi^{-}\bar{\Xi}^{+} \to
\Lambda\bar{\Lambda}\pi^{+}\pi^{-} \to
p\bar{p}\pi^{+}\pi^{+}\pi^{-}\pi^{-}$ using a missing track
method. Events are selected requiring all the tracks to be
reconstructed except the one to be studied, and the invariant mass of
the missing track predicted from the reconstructed tracks must be
consistent with the invariant mass of the track to be studied. The
tracking efficiency is then the fraction of the selected events with
at least one additional track. The PID efficiency is obtained
via the same missing track method. The tracking efficiency for MC
simulated events is found to agree with that determined using data
within $2.0\%$ for each $p$, $\bar{p}$ track and $1.0\%$ for each
$\pi^{+}$ and $\pi^{-}$ track.  Adding the uncertainties from $p$,
$\bar{p}$, $\pi^{+}$s and $\pi^{-}$s in quadrature, $4.0\%$ is taken
as the systematic uncertainty for the six charged track final states.
The PID efficiency for MC simulated events agrees with that determined
using the data within $1.0\%$ for each $p$, $2.0\%$ for each $\bar{p}$
and $0.5\%$ for each $\pi^{+}$ and $\pi^{-}$, so 2.6\% is taken as the
systematic uncertainty for the $p\bar{p}\pi^{+}\pi^{+}\pi^{-}\pi^{-}$
identification by adding the uncertainties in quadrature.

The photon reconstruction efficiency is studied via three different
methods: the missing photon method, the missing $\pi^{0}$ method and
the $\pi^{0}$ decay angle method with $\psi^{\prime}\to\pi^{+}\pi^{-}
J/\psi \to \pi^{+}\pi^{-}\rho^{0}\pi^{0}$,
$\psi^{\prime}\to\pi^{0}\pi^{0}J/\psi \to \pi^{0}\pi^{0}l^{+}l^{-}$
and $J/\psi \to \rho^{0}\pi^{0}$ events, respectively. The efficiency
difference between data and MC simulated events is within $1.0\%$ for
each photon~\cite{HYP_PHOTON}. Thus, $5.0\%$ and $1.0\%$ are taken as
the systematic uncertainty due to photon reconstruction for
$\eta_{c} \to \Sigma^{+}\bar{\Sigma}^{-}$ and $\eta_{c} \to
\Xi^{-}\bar{\Xi}^{+}$, whose final states contain five photons and one
photon, respectively.

The uncertainty of the $\pi^{0}$ selection is determined with the data
sample $J/\psi \to \bar{\Sigma}^{-} \Lambda\pi^{+}\to
\pi^{0}p\bar{p}\pi^{+}\pi^{-}$. The $\pi^{0}$ selection efficiency is
determined from the change in the $\bar{\Sigma}^{-}$ signal yield from
fitting the $\Lambda\pi^{+}$ recoiling mass distribution with and
without the $\pi^{0}$ selection requirement.  The difference between
beam data and MC simulated events on the $\pi^{0}$-selection
efficiency is within $0.5\%$ per $\pi^{0}$; hence $1.0\%$ is taken as
the systematic uncertainty from $\pi^{0}$ selection for
$\eta_{c} \to \Sigma^{+}\bar{\Sigma}^{-}$.

Samples of $J/\psi \to \gamma K^{*+}\bar{K}^{*-} \to \gamma
K^{+}K^{-}\pi^{0}\pi^{0}$, $J/\psi \to p\bar{p}\eta \to
p\bar{p}\pi^{0}\pi^{0}\pi^{0}$ and $J/\psi \to \gamma \eta_{c}\to
\gamma K^{+}K^{-}\pi^{+}\pi^{+}\pi^{-}\pi^{-}$ are selected to study
the efficiency difference between beam data and simulated MC events in
the kinematic fitting analysis for $\eta_{c} \to
\Sigma^{+}\bar{\Sigma}^{-}$ and $\eta_{c} \to
\Xi^{-}\bar{\Xi}^{+}$. In
$\eta_{c}\to\Sigma^{+}\bar{\Sigma}^{-}$, the sample of $J/\psi \to
\gamma K^{*+}\bar{K}^{*-} \to \gamma K^{+}K^{-}\pi^{0}\pi^{0}$ is
selected to estimate the efficiency of the first two $\chi^{2}_{4C}$
requirements: $\chi^{2}_{4C}(p\bar{p}\pi^{0}\pi^{0}) > 200$ and
$\chi^{2}_{4C}(\gamma p\bar{p}\pi^{0}\pi^{0}) <
\chi^{2}_{4C}(\gamma\gamma p\bar{p}\pi^{0}\pi^{0})$, and the
efficiency of the $\chi^{2}_{4C}(\gamma p\bar{p}\pi^{0}\pi^{0}) < 30$
requirement is estimated by the change in the $\eta$ signal yield from
fitting the $p\bar{p}$ recoiling mass distribution from
$J/\psi \to p\bar{p}\eta \to p\bar{p}\pi^{0}\pi^{0}\pi^{0}$ when the
$\chi^{2}_{4C}$ of the $J/\psi \to p\bar{p}\pi^{0}\pi^{0}\pi^{0}$
hypothesis is less than $30$. In $\eta_{c} \to
\Xi^{-}\bar{\Xi}^{+}$, we select a clean $J/\psi \to \gamma
\eta_{c}\to \gamma K^{+}K^{-}\pi^{+}\pi^{+}\pi^{-}\pi^{-}$ sample,
plot the 4C kinematic fitting efficiency at different $\chi^{2}_{4C}$
requirements and obtain the efficiency for the requirements as
described in the event selection section. The estimated systematic
uncertainties are $4.3\%$ and $3.8\%$ from kinematic fitting for
$\eta_{c} \to \Sigma^{+}\bar{\Sigma}^{-}$ and $\eta_{c} \to
\Xi^{-}\bar{\Xi}^{+}$, respectively.

The uncertainty from the $\Sigma^{+}$-mass window requirement is estimated by selecting a sample of
 $J/\psi \to \Sigma^{+}\bar{\Sigma}^{-}$ events and by studying the efficiency difference
 between beam data and simulated MC events. An uncertainty of $0.6\%$ is found.

The uncertainties from the vertex fits and from the $\Xi^{-}$, $\Lambda$-mass window
 requirements are estimated from a sample of $J/\psi \to \Xi^{-}\bar{\Xi}^{+} \to \Lambda\bar{\Lambda}\pi^{+}\pi^{-} \to p\bar{p}\pi^{+}\pi^{+}\pi^{-}\pi^{-}$ events. The efficiency difference between beam data and simulated MC
 events is within $0.6\%$, $0.3\%$ and $0.3\%$ for the vertex fits, $\Xi^{-}$
and $\Lambda$-mass window requirements, respectively.

Uncertainties from event generators are studied by comparing results with different models that were used for the
generation of the signal events. The decays $\eta_{c} \to \Sigma^{+}\bar{\Sigma}^{-}$
and $\eta_{c} \to \Xi^{-}\bar{\Xi}^{+}$ are generated with another model using the helicity amplitude,
and assuming that the baryons are uniformly distributed in the rest frame of $\eta_{c}$; the decays
$\Sigma^{+} \to p\pi^{0}$, $\Xi^{-} \to \Lambda\pi^{-}$ and $\Lambda \to p\pi^{-}$ are generated
with another model, which takes parity violation effects into consideration.
The efficiency differences are $0.4\%$ and $2.8\%$ for $\eta_{c} \to \Sigma^{+}\bar{\Sigma}^{-}$
and $\eta_{c} \to \Xi^{-}\bar{\Xi}^{+}$, respectively.

Uncertainties from fitting the invariant-mass distributions of $\Sigma^{+}\bar{\Sigma}^{-}$
and $\Xi^{-}\bar{\Xi}^{+}$ pairs are estimated by varying signal and background shapes and the
 corresponding fitting range. The mass and width of the $\eta_{c}$ are varied
 by $1\sigma$ according to the new measurements from BESIII~\cite{BESIII_ETAC};
 the damping function is changed from the form used by KEDR~\cite{KEDR} to
 $e^{-\frac{E_{\gamma}^{2}}{8\beta^{2}}}$ with $\beta$ fixed at $65$~MeV, which was used
 by CLEO~\cite{CLEO_DAMPING}; the MC signal shape is convoluted with a Gaussian with the width
 as a free parameter in the fit to study a possible uncertainty from the mass resolution determined
 from simulated MC events; the background shapes are varied either through the order of the polynomial or the
normalization of fixed parts; the fitting range is varied to either a narrower or a wider one.
 Taking all the factors described above into account and by adding the uncertainties from each
factor in quadrature, the uncertainties due to the fitting procedures are estimated
 to be $4.7\%$ and $6.4\%$ for $\eta_{c} \to \Sigma^{+} \bar{\Sigma}^{-}$
and $\eta_{c} \to \Xi^{-}\bar{\Xi}^{+}$, respectively.

The measured branching fractions of the peaking background channels have uncertainties
around $\sim 20-30\%$. The uncertainties from the number of peaking background events
 are estimated by assigning conservative estimates of $50\%$ to the uncertainties of the measured
branching fractions of $\eta_{c} \to p\bar{p} \pi^{0}\pi^{0}$, $\eta_{c} \to
 p\bar{p}\pi^{+}\pi^{+}\pi^{-}\pi^{-}$ and $\eta_{c} \to \Lambda\bar{\Lambda}\pi^{+}\pi^{-}$.

The total number of $J/\psi$ events is determined from analyzing $J/\psi$ inclusive hadronic
decays, and the uncertainty is $1.2\%$~\cite{JPSI_TTNM}.

Limited knowledge of the branching fractions, $\mathcal{B}(J/\psi \to \gamma \eta_{c})$,
$\mathcal{B}(\Sigma^{+}\to p\pi^{0})$, and $\mathcal{B}(\Lambda\to p\pi^{-})$ contribute
 23.5\%, 0.6\%, and 0.8\% uncertainty to the measured branching fractions, respectively~\cite{PDG}.
The first of these is the dominant source of systematic uncertainty, as indicated in Table~\ref{tab:SE_SIGMA_XI}.

All the systematic uncertainties and their sources for the channels $\eta_{c} \to \Sigma^{+}\bar{\Sigma}^{-}$ and $\eta_{c} \to \Xi^{-}\bar{\Xi}^{+}$ are summarized in Table~\ref{tab:SE_SIGMA_XI}.
The quadratic sum of all the systematic uncertainties that solely stem from our experiment are $8.7$\% and $9.5$\% in the branching fraction measurements of $\eta_{c} \to \Sigma^{+}\bar{\Sigma}^{-}$ and $\eta_{c} \to \Xi^{-}\bar{\Xi}^{+}$,
 respectively. The total systematic uncertainty is about $25$\% for both measurements.

\begin{table}[h!]
\begin{center}
\caption{Systematic uncertainties (\%) in the branching fraction measurements of $\eta_{c} \to \Sigma^{+}\bar{\Sigma}^{-}$
and $\eta_{c} \to \Xi^{-}\bar{\Xi}^{+}$.}

\begin{tabular}{l|c|c}
\hline \hline
 Source                &   $\eta_{c} \to \Sigma^{+}\bar{\Sigma}^{-}$     & $\eta_{c} \to \Xi^{-}\bar{\Xi}^{+}$        \\
\hline
Tracking and PID       &       $2.8$                                     &          $4.8$                             \\
Photon reconstruction  &       $5.0$                                     &          $1.0$                             \\
$\pi^{0}$ selection    &       $1.0$                                     &             -                              \\
$\Sigma^{+}$ mass window  &    $0.6$                                     &             -                              \\
$\Lambda$ mass window  &         -                                       &          $0.3$                             \\
$\Xi^{-}$ mass window  &         -                                       &          $0.3$                             \\
Vertex fits            &         -                                       &          $0.5$                             \\
Kinematic fits         &       $4.3$                                     &          $3.8$                             \\
Signal fitting         &       $4.7$                                     &          $6.4$                             \\
Event generators       &       $0.4$                                     &          $2.8$                             \\
Peaking background     &       $0.3$                                     &          $1.3$                             \\
$N_{J/\psi}$           &       $1.2$                                     &          $1.2$                             \\
Intermediate states    &       $23.5$                                    &          $23.6$                            \\
\hline
Total (BESIII)         &       $8.7$                                     &          $9.5$                             \\
\hline
Total                  &       $25.1$                                    &          $25.5$                            \\
\hline \hline
\end{tabular}
\label{tab:SE_SIGMA_XI}
\end{center}
\end{table}

\section{\boldmath{Summary}}\label{Summary}
Using $2.25\times10^{8}$ $J/\psi$ events collected with the BESIII
detector, the decays $J/\psi \to \gamma\eta_{c} \to
\gamma\Sigma^{+}\bar{\Sigma}^{-}$ and $J/\psi \to \gamma\eta_{c} \to
\gamma\Xi^{-}\bar{\Xi}^{+}$ are observed for the first time, and their
branching fractions are measured to be:
\begin{center}
 $\mathcal{B}(J/\psi \to \gamma \eta_{c} \to \gamma \Sigma^{+}\bar{\Sigma}^{-})=(3.60\pm0.48\pm0.31)\times10^{-5}$, \\
 $\mathcal{B}(J/\psi \to \gamma \eta_{c} \to \gamma \Xi^{-}\bar{\Xi}^{+})=(1.51\pm0.27\pm0.14)\times10^{-5}$.
\end{center}

\noindent Using the known value of $\mathcal{B}(J/\psi \to \gamma
\eta_{c})=(1.7\pm0.4)\%$~\cite{PDG}, the branching fractions
of $\eta_{c} \to \Sigma^{+}\bar{\Sigma}^{-}$ and $\eta_{c} \to
\Xi^{-}\bar{\Xi}^{+}$ are obtained:
\begin{center}
$\mathcal{B}(\eta_{c} \to \Sigma^{+}\bar{\Sigma}^{-})=(2.11\pm0.28\pm0.18 \pm0.50)\times10^{-3}$, \\
$\mathcal{B}(\eta_{c} \to \Xi^{-}\bar{\Xi}^{+})=(0.89\pm0.16\pm0.08 \pm0.21)\times10^{-3}$,       \\
\end{center}
where the first uncertainties are statistical, the second systematic,
and the third uncertainties are from the precision of the intermediate branching
fractions.

Table~\ref{tab:result} compares the results of our measurements with
the predictions from charmed-meson loop calculations~\cite{XHL_JPG}.
The measured branching fraction of $\eta_{c} \to
\Sigma^{+}\bar{\Sigma}^{-}$ is larger than the prediction, while the
measured branching fraction of $\eta_{c} \to \Xi^{-}\bar{\Xi}^{+}$
agrees with the prediction. Among the four $\eta_{c}$ baryonic decays
($\eta_{c} \to
p\bar{p}$,~$\Lambda\bar{\Lambda}$,~$\Sigma^{+}\bar{\Sigma}^{-}$, and
$\Xi^{-}\bar{\Xi}^{+}$), only $\eta_{c} \to
\Sigma^{+}\bar{\Sigma}^{-}$ disagrees with the prediction, which may
indicate the violation of SU(3) symmetry.

The precision of the branching fraction measurements of $\eta_{c} \to
\Sigma^{+}\bar{\Sigma}^{-}$ and $\eta_{c} \to \Xi^{-}\bar{\Xi}^{+}$
are limited by statistics, and the dominating systematic error stems
from the uncertainty in the branching fraction of $J/\psi \to
\gamma\eta_{c}$, which cannot be reduced without a thorough
theoretical understanding of the $\eta_{c}$ line shape in M1
transitions in the charmonium system.

\section {\boldmath{Acknowledgments}}

  The BESIII collaboration thanks the staff of BEPCII and the computing center for their hard efforts.
  This work is supported in part by the Ministry of Science and Technology of China under Contract No. 2009CB825200;
  National Natural Science Foundation of China (NSFC) under Contracts Nos. 10625524, 10821063, 10825524, 10835001, 10935007, 11125525, 10979038, 11079030, 11005109, 11179007, 11275189; Joint Funds of the National Natural Science Foundation of China under Contracts Nos. 11079008, 11179007; the Chinese Academy of Sciences (CAS) Large-Scale Scientific Facility Program; CAS under Contracts Nos. KJCX2-YW-N29, KJCX2-YW-N45; 100 Talents Program of CAS; Research Fund for the Doctoral Program of Higher Education of China under Contract No. 20093402120022; Istituto Nazionale di Fisica Nucleare, Italy; Ministry of Development of Turkey under Contract No. DPT2006K-120470; U. S. Department of Energy under Contracts Nos. DE-FG02-04ER41291, DE-FG02-91ER40682, DE-FG02-94ER40823; U.S. National Science Foundation; University of Groningen (RuG); the Helmholtzzentrum fuer Schwerionenforschung GmbH (GSI), Darmstadt; and WCU Program of National Research Foundation of Korea under Contract No. R32-2008-000-10155-0.

\end{document}